\documentclass[conference]{IEEEtran}
\IEEEoverridecommandlockouts
\usepackage{cite}
\usepackage{amsmath,amssymb,amsfonts}
\usepackage{graphicx}
\usepackage{textcomp}
\usepackage{xcolor}
\usepackage{booktabs}
\usepackage{algorithm}
\usepackage{algpseudocode}
\usepackage{array}
\usepackage{multirow}
\usepackage{url}
\usepackage{tikz}
\usetikzlibrary{arrows.meta, positioning, shapes.geometric, calc, fit}

\def\BibTeX{{\rm B\kern-.05em{\sc i\kern-.025em b}\kern-.08em
    T\kern-.1667em\lower.7ex\hbox{E}\kern-.125emX}}

\begin{document}
\title{FAIRGate: Agentic Runtime Policy Selection and Verified Repair for Fairer Recommendation Delivery%
\thanks{\copyright~2026 IEEE. Personal use of this material is permitted. Permission from IEEE must be obtained for all other uses, in any current or future media, including reprinting/republishing this material for advertising or promotional purposes, creating new collective works, for resale or redistribution to servers or lists, or reuse of any copyrighted component of this work in other works.
Accepted for publication at the 2026 IEEE International Conference on Responsible Artificial Intelligence (IRAI), Melbourne, Australia.
}}

\author{
\IEEEauthorblockN{Navkar Jain, Enock Ayiku, Kenneth Fletcher}
\IEEEauthorblockA{
Department of Computer Science, University of Massachusetts Boston\\
Email: \{n.jain001, enock.ayiku001, kenneth.fletcher\}@umb.edu
}
}

\maketitle

\begin{abstract}
Fairness-aware recommender systems often encode one fairness objective during training or apply a fixed re-ranking rule after scoring. In deployment, however, the appropriate fairness action can vary across requests because the candidate pool, item metadata, exposure debt, user preference history, and utility budget all change at runtime. We propose FAIRGate, a model-agnostic delivery controller that places an agentic policy-selection layer between a base ranker and the final top-$K$ list. The agent does not define fairness objectives from scratch. It chooses from a predefined policy library covering long-tail exposure, item-group exposure balance, concentration control, calibration protection, and abstention. A deterministic executor then applies the selected policy through constrained repair, and a validator accepts the result only when relevance, feasibility, metadata, calibration, and audit constraints are satisfied. The final language-model component receives only the verified audit trace and produces evidence-linked narration after the ranking decision is fixed. This design gives the agent runtime responsibility while keeping fairness enforcement measurable, bounded, and replayable.
\end{abstract}


\section{Introduction}
Recommender systems decide which items appear in a user's top-$K$ list. This decision shapes user choice and affects the visibility received by long-tail items, authors, creators, genres, and other item groups. Fairness interventions commonly enter the pipeline through training objectives, post-processing re-rankers, or explanations. These tools are important, but they often assume a fixed fairness action. In practice, one fixed action can be too rigid. A user with a diverse candidate pool may support long-tail repair, while another user with sparse history may need calibration protection or abstention because feasible alternatives are unavailable.

This paper studies fairness as a runtime delivery-control problem. Given a base ranker output, the system must decide which fairness policy applies now, which repair is feasible, and whether the repaired list is ready to deliver. A purely deterministic priority order is reproducible, but it can treat distinct delivery contexts in the same way. A fully unconstrained language-model agent is flexible, but it can make unstable or unsupported fairness judgments. FAIRGate takes a bounded path: the agent chooses and plans from a fixed policy library using structured metrics, while deterministic validation enforces the selected policy before delivery.

The design rests on a separation of roles. The base recommender estimates relevance. The policy-selection agent identifies the current fairness context and chooses applicable policies. The repair-planning agent proposes bounded actions. The deterministic executor and validator apply and verify those actions. The narration agent communicates only the verified trace. This makes the system agentic at the control layer, not arbitrary at the fairness layer.

This paper makes three contributions.
\begin{itemize}
    \item [1.] FAIRGate formalizes bounded agentic fairness control for recommendation delivery, where an agent selects among predefined exposure and calibration policies rather than inventing fairness criteria at inference time.
    \item [2.] We introduce a verified repair loop that executes agent-selected policies only when deterministic relevance, feasibility, metadata, calibration, and audit checks are satisfied.
    \item [3.] We define an audit-trace contract that converts policy selections, repair actions, validator outcomes, and abstention decisions into typed evidence keys for claim-level explanation verification.
\end{itemize}
Together, these contributions move recommendation delivery closer to fairer practice by combining adaptive policy choice with enforceable safeguards.

\section{Related Work and Design Gap}
\textbf{Fair exposure and popularity bias.} Fair ranking work formalizes exposure as position-weighted visibility and studies how exposure should relate to merit or group membership \cite{singh2018fairness,biega2018equity}. Recommender systems add another challenge because repeated exposure to popular items can reduce catalog coverage and long-tail discovery. Abdollahpouri et al. study popularity-bias mitigation and metrics such as average recommendation popularity, average percentage of long-tail items, and average coverage of long-tail items \cite{abdollahpouri2019popularity}. Provider and item-side fairness methods further study how visibility is distributed across creators, providers, and item groups \cite{burke2017multisided,xu2023pmmf}. FAIRGate builds on this literature by treating exposure objectives as deployable policies that must be selected and validated for each delivery context.

\textbf{Multi-stakeholder and calibrated recommendation.} Recommendation fairness is not only a user-accuracy problem. It includes user utility, provider visibility, category representation, calibration, and the distribution of gains and losses across users \cite{yao2017beyond,deldjoo2022fairness,ekstrand2023distributionally}. Calibration is important for runtime repair because exposure improvement can move the delivered list away from a user's historical content distribution. FAIRGate therefore includes calibration protection as a selectable policy and as a validator condition, rather than treating every exposure repair as automatically deliverable.

\textbf{LLMs and agentic recommendation.} LLMs have been used as zero-shot rankers, user simulators, and agentic components for recommender systems \cite{hou2024llmrank,wang2023recagent,zhang2024agentcf}. These systems show the potential of language agents to reason over recommendation context, but they also raise governance questions when an LLM directly controls ranking or evaluation. LLM-as-judge studies report sensitivity to prompts, positions, and evaluation framing \cite{zheng2023judging,huang2024judge}. FAIRGate uses LLMs for policy selection, repair planning, and narration, but constrains their authority through a policy library, deterministic execution, and post-hoc validation.

\textbf{Design gap.} Existing work gives fair ranking objectives, re-ranking algorithms, and LLM-based recommendation agents. Less work studies the final delivery decision as a bounded agentic control point. The gap is not only whether a list can be made fairer. It is whether the system can decide which fairness policy is relevant for the current request, apply it without exceeding a relevance budget, and explain the verified action without unsupported claims. FAIRGate targets this gap by moving adaptivity to policy selection while keeping enforcement deterministic.

\section{FAIRGate}
\subsection{Research Questions}
We evaluate FAIRGate through four questions.
\begin{itemize}
    \item \textit{RQ1:} Does runtime policy selection improve long-tail and item-group exposure satisfaction compared with base rankers, fixed re-rankers, and deterministic fixed-policy gates under the same utility budget?
    \item \textit{RQ2:} Does validator-constrained repair preserve recommendation utility while improving selected exposure policies?
    \item \textit{RQ3:} Does the agent choose different policies for different runtime contexts, and do those choices explain gains over fixed policy orderings?
    \item \textit{RQ4:} Does audit-grounded narration reduce unsupported fairness claims compared with direct LLM explanations?
\end{itemize}

\subsection{Problem Setup and Policy Semantics}
For a user $u$, a base ranker returns a candidate pool $C_u$ with relevance scores $s_u(i)$ for item $i$. The system must deliver a list $R_u$ of length $K$. Each item has metadata, such as popularity tier, genre or tag group, and author or item-group label. Let $v(r)$ be the visibility discount at rank $r$. The exposure received by group $g$ in a delivered list is
\begin{equation}
E_g(R_u)=\sum_{i\in R_u} v(\mathrm{rank}_{R_u}(i))\mathbb{1}[g(i)=g].
\end{equation}
A policy $P$ is a tuple $(q, c, a, \nu)$, where $q$ is a measurable objective, $c$ is a set of constraints, $a$ is an allowed repair action set, and $\nu$ is a validator. The policy library includes $P_T$ for long-tail floors, $P_G$ for item-group exposure balance, $P_C$ for concentration caps, $P_{Cal}$ for calibration protection, and $P_A$ for abstention when metadata or feasibility is insufficient. The agent selects a policy subset $A_u\subseteq\mathcal{P}$ from this library using runtime state $x_u$, but the validator decides whether the resulting repair is deliverable.

In the current implementation, group labels come from available item metadata, such as genre or popularity tier. They are used to evaluate exposure and calibration behavior in a controlled recommendation setting. FAIRGate is therefore a delivery-control framework over explicitly defined policy targets, not a claim that any single metadata grouping exhausts fairness in a deployment domain.

\subsection{System Architecture}
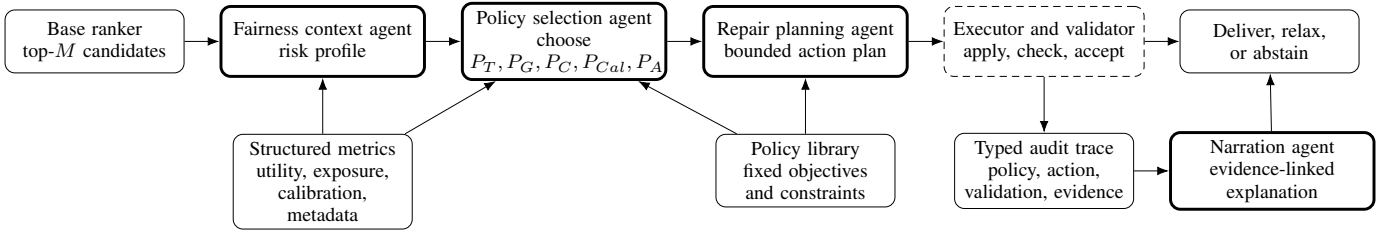
\begin{figure*}[t]
\centering
\resizebox{\textwidth}{!}{%
\begin{tikzpicture}[
    font=\footnotesize,
    >=Latex,
    box/.style={draw, rounded corners, align=center, minimum height=9mm, text width=2.35cm, inner sep=3pt},
    agent/.style={draw, rounded corners, align=center, minimum height=10mm, text width=2.7cm, inner sep=3pt, very thick},
    val/.style={draw, rounded corners, align=center, minimum height=10mm, text width=2.65cm, inner sep=3pt, densely dashed},
    node distance=5mm and 5mm
]
\node[box] (ranker) {Base ranker\\top-$M$ candidates};
\node[agent, right=of ranker] (context) {Fairness context agent\\risk profile};
\node[agent, right=of context] (select) {Policy selection agent\\choose $P_T,P_G,P_C,P_{Cal},P_A$};
\node[agent, right=of select] (plan) {Repair planning agent\\bounded action plan};
\node[val, right=of plan] (exec) {Executor and validator\\apply, check, accept};
\node[box, right=of exec] (deliver) {Deliver, relax,\\or abstain};

\node[box, below=8mm of context] (metrics) {Structured metrics\\utility, exposure, calibration, metadata};
\node[box, below=8mm of plan] (library) {Policy library\\fixed objectives and constraints};
\node[box, below=8mm of exec] (audit) {Typed audit trace\\policy, action, validation, evidence};
\node[agent, right=of audit] (narrate) {Narration agent\\evidence-linked explanation};

\draw[->] (ranker) -- (context);
\draw[->] (context) -- (select);
\draw[->] (select) -- (plan);
\draw[->] (plan) -- (exec);
\draw[->] (exec) -- (deliver);
\draw[->] (metrics) -- (context);
\draw[->] (metrics) -- (select);
\draw[->] (library) -- (select);
\draw[->] (library) -- (plan);
\draw[->] (exec) -- (audit);
\draw[->] (audit) -- (narrate);
\draw[->] (narrate) -- (deliver);
\end{tikzpicture}%
}
\caption{FAIRGate architecture. Agents interpret the runtime recommendation context, select policies, and plan repair. Deterministic execution and validation keep final delivery within the predefined policy library and utility budget.}
\label{fig:architecture}
\end{figure*}

Fig.~\ref{fig:architecture} shows the architecture. The important distinction is between agentic choice and deterministic authority. Agents select policies and propose repair plans. Validators enforce feasibility, relevance, calibration, metadata coverage, and traceability before the list reaches the user.

\subsection{Runtime Agents}
\textbf{Fairness context agent.} This agent receives the base top-$K$ list, candidate-pool summaries, popularity tiers, group exposure, calibration divergence, metadata coverage, and relevance margins. It outputs a structured profile: popularity concentration, group exposure debt, calibration sensitivity, feasible replacement count, and abstention risk. The profile is JSON-like and contains no free-form fairness verdict.

\textbf{Policy selection agent.} This agent selects one or more policies from the fixed library. For example, it may choose $P_T+P_{Cal}$ when the base list has low long-tail coverage but enough relevant tail candidates, or $P_A$ when metadata coverage is below the minimum threshold. The agent must cite metric triggers for every selected policy. It cannot create new fairness goals.

\textbf{Repair planning agent.} This agent proposes a bounded plan, such as moving eligible long-tail items upward, limiting author repetition, or preserving genre calibration. The plan specifies allowed swaps, relevance floors, policy priority, and fallback behavior. The executor converts the plan into candidate list edits.

\textbf{Validator and narration agent.} The validator checks whether the proposed repair satisfies the selected policy, preserves utility within $\epsilon$, keeps calibration within $\tau_C$, respects metadata coverage, and records every item movement. The narration agent receives only this trace and produces user-facing text with evidence keys. This boundary prevents language fluency from becoming ranking authority.

\subsection{Agent Input and Output Contract}
The agentic layer is constrained by a typed input-output contract. Each agent receives only precomputed metrics, candidate summaries, and policy descriptors. Each agent returns a structured object that can be parsed and checked. The contract is designed to make runtime choice adaptive without allowing unbounded fairness invention. Table~\ref{tab:iocontract} summarizes the contract used in the experiments.

\begin{table*}[t]
\caption{Agent input-output contract. The contract makes the agentic step inspectable and gives the validator a fixed schema to check.}
\centering
\scriptsize
\setlength{\tabcolsep}{2.5pt}
\begin{tabular}{p{0.15\linewidth}p{0.27\linewidth}p{0.27\linewidth}p{0.22\linewidth}}
\toprule
\textbf{Module} & \textbf{Inputs} & \textbf{Allowed outputs} & \textbf{Validator checks} \\
\midrule
Context agent & Base top-$K$, top-$M$ candidate summaries, exposure metrics, metadata coverage, calibration drift & Risk profile with metric-backed labels for tail risk, group debt, concentration, calibration sensitivity, and feasibility & Every risk label cites an input metric and threshold \\
Policy agent & Risk profile, fixed policy library, validation-selected thresholds & Policy subset $A_u$, priority order, and metric triggers & Selected policies must belong to $\mathcal{P}$ and match at least one trigger \\
Repair planner & Policy subset, candidate pool, relevance margins, allowed action set & Swap plan, protected utility floor, fallback path, and relaxation priority & Every item movement must be feasible and score-bounded \\
Executor & Repair plan and ranked candidates & Candidate edits and post-gate metrics & Utility loss, JSD drift, metadata coverage, and policy pass are recomputed \\
Narration agent & Typed audit trace only & Explanation claims linked to evidence keys & Unsupported claims are rejected or rewritten \\
\bottomrule
\end{tabular}
\label{tab:iocontract}
\end{table*}

A policy-selection prompt therefore follows a narrow form: given the metrics and the fixed library, choose the smallest policy subset that addresses the dominant risk while protecting the utility and calibration budgets. The prompt does not ask whether the list is fair. It asks which listed policy applies, which metric triggered it, and which validator must confirm it. This makes the agent a runtime controller over explicit choices, not a free-form moral judge.

\subsection{Verified Repair Objective}
The executor scores a feasible replacement item $i$ for user $u$ using
\begin{equation}
S(i,u)=\alpha s_u(i)+\beta d_T(i)+\gamma d_G(i)-\lambda \Delta Cal(i,u),
\end{equation}
where $d_T(i)$ is the long-tail gain, $d_G(i)$ is group exposure-debt reduction, and $\Delta Cal(i,u)$ is the predicted calibration drift. The coefficients $\alpha$, $\beta$, $\gamma$, and $\lambda$ are non-negative validation-selected weights that control the trade-off among relevance, tail exposure, group-debt reduction, and calibration protection. A repair is accepted only if it improves the selected policy objective and keeps utility loss below $\epsilon$:
\begin{equation}
\frac{U(R_u)-U(R'_u)}{U(R_u)}\leq \epsilon.
\end{equation}
The final decision is \textit{deliver} when validation passes, \textit{relax} when a lower-priority policy must be softened to protect utility or calibration, and \textit{abstain} when metadata or feasible replacement conditions fail.

\begin{algorithm}[t]
\caption{FAIRGate Delivery Procedure}
\label{alg:fairgate}
\begin{algorithmic}[1]
\Require candidate pool $C_u$, scores $s_u$, policy library $\mathcal{P}$, utility budget $\epsilon$
\State compute runtime state $x_u$ from utility, exposure, metadata, and calibration metrics
\State context agent outputs structured risk profile $z_u$
\State policy agent selects $A_u\subseteq\mathcal{P}$ and cites metric triggers
\If{$P_A\in A_u$ because feasibility is insufficient}
    \State abstain with typed audit reason
\EndIf
\State repair agent proposes bounded plan $\pi_u$ using $A_u$
\State executor applies feasible edits to produce $R'_u$
\State validator checks relevance, policy, calibration, metadata, and trace constraints
\If{all checks pass}
    \State deliver $R'_u$ and store audit trace
\ElsIf{lower-priority relaxation preserves the selected primary policy}
    \State relax, revalidate, and store relaxation reason
\Else
    \State abstain and store infeasibility reason
\EndIf
\State narration agent generates explanation only from validated audit keys
\State \Return decision, delivered list if any, audit trace, explanation
\end{algorithmic}
\end{algorithm}

\begin{table}[t]
\caption{Policy library used by the policy-selection agent. Policies are fixed before evaluation. The agent selects among them at runtime.}
\centering
\scriptsize
\setlength{\tabcolsep}{3pt}
\begin{tabular}{p{0.13\linewidth}p{0.28\linewidth}p{0.43\linewidth}}
\toprule
\textbf{Policy} & \textbf{Trigger signal} & \textbf{Validated action} \\
\midrule
$P_T$ & Low long-tail share & Add feasible tail items within relevance floor \\
$P_G$ & Exposure debt by author, genre, tag, or group & Prioritize under-exposed groups in replacement set \\
$P_C$ & High group or author concentration & Limit repeated exposure from dominant group \\
$P_{Cal}$ & High predicted calibration drift & Reject repairs that move JSD above $\tau_C$ \\
$P_A$ & Metadata gaps or no feasible repairs & Abstain or deliver base list with audit reason \\
\bottomrule
\end{tabular}
\label{tab:policylibrary}
\end{table}

\section{RQ-Driven Evaluation Design}
\subsection{Datasets and Base Rankers}
We evaluate FAIRGate on MovieLens 1M. The dataset contains approximately one million ratings with movie genres, timestamps, and user attributes \cite{harper2015movielens}. Ratings of 4 or 5 are treated as positive implicit feedback. We use timestamp-based leave-last-one-out splitting, with each user's latest interaction used for test and the second-latest interaction used for validation. Popularity tiers, genre-group counts, and calibration targets are computed from training data only.

MovieLens 1M supports a controlled evaluation of long-tail movie exposure, genre-group exposure, and calibration. Base rankers include TopPop, item-kNN, and BPR-MF \cite{rendle2009bpr}. RecBole standardizes training and evaluation \cite{zhao2021recbole}. FAIRGate operates on the top-$M=100$ candidate pool and delivers $K=10$ items.

\subsection{Baselines, Metrics, and Ablations}
The main baselines are the base BPR-MF ranker, a static long-tail re-ranker, a deterministic fixed-order FAIRGate, an agentic variant without validation, and the full agentic FAIRGate. The key comparison is deterministic fixed-order FAIRGate versus full agentic FAIRGate because it isolates the value of runtime policy selection under the same policy library and utility budget.

Utility metrics are NDCG@10, Recall@10, and relative NDCG loss. Exposure metrics are LT-Cov@10, exposure Gini, and policy-pass rate. Calibration uses Jensen-Shannon divergence between the user's historical genre distribution and the delivered list. Agent metrics are validator rejection rate and selected-policy agreement with a validation-derived oracle policy label. Narration metrics are unsupported claim rate, evidence coverage, audit-faithfulness rate, and sensitive-attribute leakage rate.

\subsection{Reproducibility Protocol}
Thresholds $\tau_T$, $\tau_G$, $\tau_C$, and $\epsilon$ are selected on validation data and frozen before test evaluation. The same user set and candidate pools are used across methods. For policy selection, the oracle label is defined as the validation policy that gives the largest exposure-policy gain while satisfying the utility budget. This label is not used to control the agent at test time. It serves only as an evaluation reference for RQ3.

Narration evaluation is separate from ranking evaluation. Each explanation is decomposed into atomic claims. A claim is supported only when it maps to a typed audit key, such as policy selection, rule trigger, item movement, validator rejection, relaxation event, abstention reason, or post-gate metric. This prevents the narration agent from receiving credit for unsupported fairness language.

\subsection{Implementation Details}
The controller is implemented as a wrapper around a ranker that returns a scored candidate pool. The base model writes one record per user request containing candidate identifiers, relevance scores, popularity tier, genre-group labels, and content tags. A metric service computes the runtime state $x_u$ before the agent is called, so the LLM does not compute exposure or calibration from raw logs.

For MovieLens 1M, the policy library includes $P_T$, $P_G$, $P_C$, $P_{Cal}$, and $P_A$, where $P_G$ is defined over genre-group exposure and $P_C$ enforces a concentration cap on the dominant group. The policy-selection step uses structured generation with schema validation and hard trigger enforcement. The agent is evaluated by whether it chooses a policy that improves the selected exposure objective under the validation-selected utility budget.

The repair executor builds a replacement set from candidates outside the base top-$K$ whose relevance scores fall within a tolerance band of the item being replaced. It then selects the swap that gives the largest improvement in the selected policy objective while passing the calibration guard. The system logs the base list, selected policy object, validator report, and final explanation for each request. These artifacts make failed requests diagnosable as metadata gaps, missing feasible replacements, utility-budget conflicts, or calibration conflicts.

\section{Experimental Results}
We run the experiments on a reproducible random sample of 1000 MovieLens 1M users. This sampling choice is made to keep LLM-call cost and latency manageable while using the same sampled users across methods. The agentic components use GroqCloud API calls with the Llama 3.1 8B model \cite{groq_llama31_8b_instant}.

\subsection{RQ1: Runtime Policy Selection and Exposure}
Table~\ref{tab:mainresults} shows that agentic FAIRGate improves the delivered exposure profile while preserving ranking quality. Compared with the base BPR-MF ranker, agentic FAIRGate increases LT-Cov@10 from 0.1055 to 0.1282, a relative gain of 21.52\%. It also gives the highest LT-Cov@10 among the static re-rank, fixed-order, and agentic variants. Its NDCG@10 is 0.0233, above the base and static re-rank baselines and close to the fixed-order gate; Recall@10 follows the same pattern.

The largest difference is policy-pass rate. Fixed-order FAIRGate passes the selected policy in 0.0910 of requests, while agentic FAIRGate reaches 0.6070. The exception is Group Gini: fixed-order FAIRGate has a lower value, 0.6581 versus 0.6672. This occurs because the fixed-order gate repeatedly applies balance and concentration repairs before considering request context, which pushes genre-group exposure closer to equality. Group Gini measures aggregate equality across item groups, while policy-pass measures whether the runtime-selected policy succeeds under relevance and calibration constraints. The agentic gate does not optimize Gini alone. It often prioritizes the currently selected policy, long-tail coverage, calibration, and relevance. Therefore, the agentic variant does not claim dominance across every fairness metric. Its advantage lies in improving selected-policy satisfaction and long-tail exposure under the same utility and calibration budget. In contrast, the fixed-order gate improves aggregate group equality more in this run, but it does so by applying the same priority sequence across requests, which reduces its ability to satisfy the policy selected for each runtime context.

\begin{table*}[t]
\caption{Main results on MovieLens 1M. Higher NDCG, Recall, LT-Cov, and Policy Pass are better. Lower Gini and JSD are better.}
\label{tab:mainresults}
\centering
\scriptsize
\setlength{\tabcolsep}{3pt}
\begin{tabular}{lccccccc}
\toprule
Method & NDCG@10 & Recall@10 & LT-Cov@10 & Group Gini & JSD-Calib & Policy Pass & Abstain \\
\midrule
Base BPR-MF & 0.0225 & 0.0530 & 0.1055 & 0.6923 & 0.1462 & -- & -- \\
Static re-rank & 0.0221 & 0.0520 & 0.1163 & 0.6901 & 0.1454 & -- & -- \\
Fixed-order FAIRGate & 0.0236 & 0.0570 & 0.1173 & 0.6581 & 0.1515 & 0.0910 & 0.0000 \\
Agentic FAIRGate & 0.0233 & 0.0560 & 0.1282 & 0.6672 & 0.1477 & 0.6070 & 0.0000 \\
\bottomrule
\end{tabular}
\end{table*}

\begin{figure}[t]
\centering
\includegraphics[width=0.9\columnwidth]{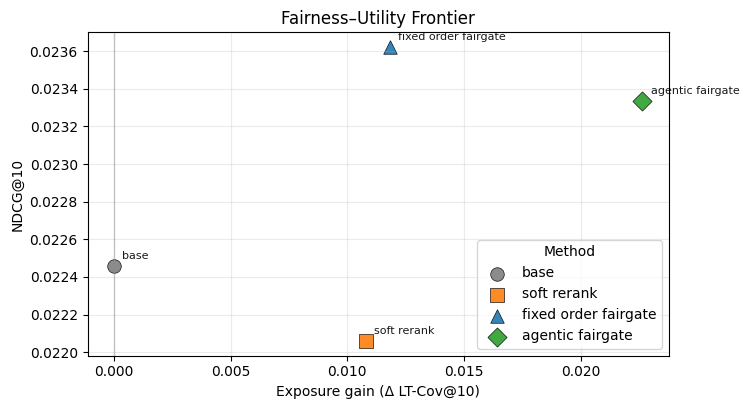}
\caption{Utility-exposure frontier on MovieLens 1M. Agentic FAIRGate achieves the highest LT-Cov@10 while keeping NDCG@10 close to the best-performing fixed-order gate.}
\label{fig:frontier}
\end{figure}

Figure~\ref{fig:frontier} visualizes the utility-exposure trade-off using NDCG@10 and LT-Cov@10. The static re-ranker improves LT-Cov@10 but lowers both NDCG@10 and Recall@10. Fixed-order FAIRGate gives the highest NDCG@10, but its LT-Cov@10 gain is smaller. Agentic FAIRGate moves farthest toward the exposure objective while retaining nearly the same NDCG@10 as the fixed-order gate. This supports the central claim of RQ1: adaptive policy selection can move recommendation delivery closer to the target exposure policy without a large relevance sacrifice.

\subsection{RQ2 and RQ3: Validator and Agent Adaptation}
Table~\ref{tab:ablation} shows that the validator changes the behavior of the agentic controller in a useful way. The no-validator and no-calibration-guard variants reach a slightly higher pass rate of 0.9290, but they do so without the final mechanism that checks whether a repair remains within the declared constraints. The full agentic gate reports a pass rate of 0.9230 with a 0.0170 rejection rate. In other words, a small share of candidate repairs is rejected rather than delivered when the validator detects a constraint issue.

The fixed policy order performs poorly in this setting. Its pass rate is 0.0000 and its NDCG loss is 0.0373, almost double the 0.0190 loss observed for the full agentic gate. This supports RQ3: the agentic controller adds value by adapting policy choice to the runtime context instead of applying a fixed policy sequence. We also observe non-trivial policy diversity in RQ3: the primary selected policy is $P_G$ in 601/1000 requests, $P_C$ in 310/1000, $P_{Cal}$ in 88/1000, and $P_T$ in 1/1000, corresponding to an empirical selection entropy of 1.2838 bits. This distribution is consistent with MovieLens 1M being dominated by group-concentration and debt-repair contexts in our setup.

Table~\ref{tab:ablation} reports validation pass among attempted repairs, while Table~\ref{tab:mainresults} reports final satisfaction of the selected policy on all evaluated requests. These metrics answer related but different questions. The ablation table measures whether proposed repairs pass the validator after a repair is attempted. The main results measure how often the delivered outcome satisfies the selected policy across the full evaluation set.

\begin{table}[t]
\caption{Ablation results for RQ2 and RQ3.}
\label{tab:ablation}
\centering
\scriptsize
\setlength{\tabcolsep}{2.5pt}
\begin{tabular}{lcccc}
\toprule
Variant & Validator Pass & NDCG Loss & Reject & Oracle Agree \\
\midrule
Full agentic gate & 0.9230 & 0.0190 & 0.0170 & 0.2890 \\
No validator & 0.9290 & 0.0190 & 0.0000 & 0.2900 \\
No calibration guard & 0.9290 & 0.0190 & 0.0000 & 0.2900 \\
Fixed policy order & 0.0000 & 0.0373 & 0.0270 & 0.4540 \\
No abstention & 0.9230 & 0.0190 & 0.0170 & 0.2890 \\
\bottomrule
\end{tabular}
\end{table}

\subsection{RQ4: Audit-Grounded Narration}
Table~\ref{tab:narration} shows a clear benefit from grounding explanations in the verified audit trace. Direct LLM explanation produces unsupported fairness claims for every evaluated explanation and gives no evidence coverage or audit-faithful claims. Audit-grounded narration reduces unsupported claims to 0.0959, increases evidence coverage to 0.9129, eliminates sensitive-attribute leakage in this run, and reaches an audit-faithfulness rate of 0.5760. These results support RQ4: the language model is more reliable when it is restricted to explaining typed policy and validator evidence rather than generating fairness rationales directly.

\begin{table}[t]
\caption{Narration results for RQ4.}
\label{tab:narration}
\centering
\scriptsize
\setlength{\tabcolsep}{2.5pt}
\begin{tabular}{lcccc}
\toprule
Method & Unsupported & Evidence & Leakage & Faithful \\
\midrule
Direct LLM explanation & 1.0000 & 0.0000 & 0.0150 & 0.0000 \\
Audit-grounded narration & 0.0959 & 0.9129 & 0.0000 & 0.5760 \\
\bottomrule
\end{tabular}
\end{table}

\subsection{Why Agentic Selection Matters}
A fixed fairness gate assumes that the same intervention is appropriate across users, candidate pools, and metadata conditions. This is restrictive in top-$K$ recommendation because feasible repair changes by request: some users have relevant long-tail candidates, others are calibration-sensitive, and some requests show concentration or metadata gaps.

Agentic FAIRGate addresses this heterogeneity at the policy-selection layer, not the enforcement layer. The agent selects among predefined policies using runtime evidence, while deterministic validation still authorizes delivery. This distinction is central to the design. It allows the system to respond differently across recommendation contexts while keeping final decisions bounded by explicit policy semantics and deterministic validation.

This also explains why Agentic FAIRGate should be compared against both static re-ranking and fixed-order gating. Static re-ranking tests whether a single exposure correction is enough. Fixed-order gating tests whether deterministic policy priority is enough. Agentic FAIRGate tests whether runtime policy choice adds value when multiple fairness actions are available but only some are appropriate for the current request. The evaluation therefore measures not only whether exposure metrics improve, but also whether policy choice, repair feasibility, and audit faithfulness work together as a delivery-control mechanism.

\section{Limitations and Future Work}
The current evaluation uses MovieLens 1M, so the results should be read as evidence from a controlled movie-recommendation setting rather than as a general claim across all domains. Movie metadata supports genre-group and popularity analysis, but it does not provide a rich provider-side marketplace structure. A natural next step is to evaluate FAIRGate on datasets with clearer creator, seller, or provider labels, including book or marketplace datasets such as Goodbooks-10k \cite{zajac2017goodbooks} when their metadata supports the selected policy targets.

The policy library also shapes the reachable fairness outcomes. FAIRGate selects among predefined policies; it does not create new fairness objectives at runtime. This design improves controllability and auditability, but it also means that missing policies must be added by system designers before deployment. Thresholds and utility budgets are selected on validation data, and different domains may require different policy definitions, metadata standards, and escalation rules.

The narration evaluation depends on claim parsing and audit-key matching. These checks are stricter than open-ended explanation grading, which is useful for auditability, but they may undercount partially supported natural-language claims. Future work can add stronger claim decomposition, human review of borderline claims, and domain-specific audit keys. Another useful direction is to study policy selection under distribution shift, where candidate-pool composition and popularity dynamics change over time.

\section{Discussion and Conclusion}
FAIRGate turns fairness delivery into a bounded runtime-control problem. The agent receives structured metrics, selects from a fixed policy library, and proposes repair plans for the current recommendation context. Deterministic validation then decides whether the repaired list is deliverable. This separation gives the agent adaptive responsibility while keeping enforcement measurable and replayable.

The results show three main findings. First, Agentic FAIRGate improves LT-Cov@10 and policy-pass rate over fixed alternatives while preserving most ranking utility. Second, validator-constrained repair prevents the agentic layer from treating every exposure gain as deliverable. Third, audit-grounded narration sharply reduces unsupported fairness claims compared with direct LLM explanation. Together, these findings support a practical path toward fairer recommendation delivery: adaptive at the policy-selection layer, deterministic at the enforcement layer, and auditable at the communication layer.

\end{document}